\documentstyle[epsfig,aps,prl,epsf,multicol]{revtex}
\draft

\begin{document}

%\twocolumn

\hsize\textwidth\columnwidth\hsize
\csname@twocolumnfalse\endcsname

\title{Depinning of a vortex chain in a disordered flow channel}

\author{R.~Besseling$^1$, T.~Dr\"{o}se$^2$, V.M. Vinokur$^3$
and P.H.~Kes$^1$}

\address{$^1$ Kamerlingh Onnes Laboratorium, Leiden
University, P.O. Box 9504, 2300 RA Leiden, the Netherlands.\\
$^2$ I. Institut f\"ur Theoretische Physik, Universit\"at Hamburg,
Jungiusstrasse 9, D-20355 Hamburg, Germany. \\$^3$ Materials
Science Division, Argonne National Laboratory, Argonne, Illinois
60439}

\date{\today}
\maketitle

\begin{abstract}
We study depinning of vortex chains in channels formed by static,
disordered vortex arrays. Depinning is governed either by the
barrier for defect nucleation or for defect motion, depending on
whether the chain periodicity is commensurate or incommensurate
with the surrounding arrays. We analyze the reduction of the gap
between these barriers as function of disorder. At large disorder,
commensurability becomes irrelevant and the pinning force is
reduced to a small fraction of the ideal shear strength of ordered
channels. Implications for experiments on channel devices are
discussed. 
\end{abstract}
\begin{multicols}{2}
\narrowtext \noindent

The depinning and dynamics of periodic elastic media in a random
potential have received a great deal of recent attention
\cite{general}. It was shown, in particular for vortex lattices
(VL's) in superconductors, that the depinning transition in most
cases involves plastic deformations \cite{plastic}. A system in
which plastic depinning can be studied in a controlled way is that
of narrow, weak pinning flow channels in a superconducting film
\cite{Pruymboom}. In such a system, strongly pinned vortices in
the channel edges (CE's) provide confinement as well as an
effective pinning potential for chains of vortices inside the
channel. By changing the magnetic field $B$, one can vary the
ratio between channel width and lattice spacing and thus induce
incommensurability between the vortex spacing inside and outside
the channel. This leads to topological defects in the channel
which sensitively affect the threshold for plastic flow, as
evidenced by oscillations of the critical current density $J_c$
versus field \cite{Pruymboom}.

Simulations of channels with perfect {\em hexagonal} vortex arrays
in the CE's, showed sharp peaks in $J_c$ for channel widths $w$
equal to an integer number $n$ of vortex row spacings, i.e.
$w=nb_0$. At mismatch ($w\neq b_0$) defects occurred in the
channel and $J_c$ vanished due to the small Peierls barrier for
defect flow \cite{RutPRL99}. This however contrasts the smooth
oscillations found experimentally. Moreover, in reality the CE
arrays may contain {\em quenched positional disorder} due to
random pinning. This should modify $J_c$ and its dependence on
commensurability. In this Letter we consider the most simple
near-to-commensurability situation $w\sim b_0$, and investigate
the effect of disorder on depinning of a {\em single vortex
chain}. We find that commensurate chains, with a periodicity equal
to that of the VL in the CE's, depin at a force $f_n$ below the
ideal shear strength $f_c^0$ by {\em nucleation of defect pairs}.
At incommensurability, $f_c$ is determined by the pinning force
$f_d$ of {\em existing defects}. At weak disorder, a gap is found
between $f_n$ and $f_d$, but for larger disorder commensurability
becomes irrelevant and $f_c$ saturates at a small fraction of
$f_c^0$. This has important consequences for the interpretation of
the experiments \cite{Pruymboom}. Generally, our results are
relevant to a wealth of 1D problems including models for interface
growth \cite{KrugPRL}, dynamics of Josephson junctions
\cite{Malomed} and charge density waves (CDW's).

\begin{figure}
\epsfig{file=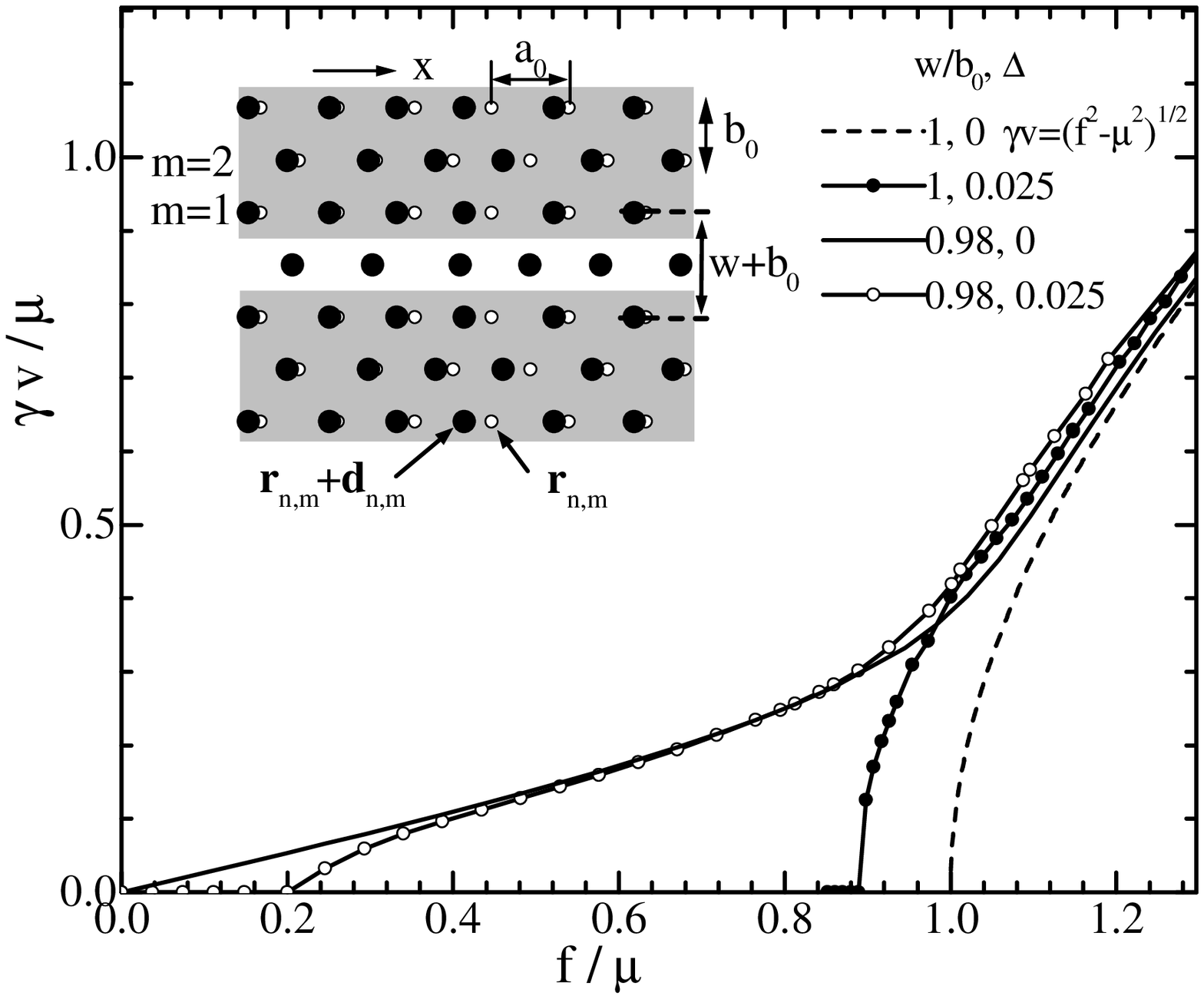, width=7cm, height=5.5cm} \vspace{0.5cm}
\caption{Simulated {\it f-v} curves at weak disorder
($\Delta=0.025$) for $w/b_0=1$ ($\bullet$) and $w/b_0=0.98$
($\circ$) (dashed and full lines are the respective results for
$\Delta=0$). Inset: channel geometry with pinned vortices in the
gray areas. Their equilibrium positions ${\bf r}_{n,m}$ are
denoted by ($\circ$). The disordered arrays (with the shifts $d_x$
exaggerated for clarity) are denoted by ($\bullet$).} \label{fv}
\end{figure}

We consider a channel at $T=0$ with boundaries (CE's) formed by
two {\it static}, disordered arrays with vortex positions ${\bf
R}_{n,m}={\bf r}_{n,m}+{\bf d}_{n,m}$, where ${\bf r}_{n,m}$
denote the hexagonal lattice with spacing $a_0=2b_0/\sqrt{3}$ (see
the inset to Fig. \ref{fv}), the channel width is defined by the
spacing between rows $m=\pm 1$ and ${\bf d}_{n,m}$ are random
shifts. We restrict ourselves to longitudinal shifts
\cite{footnote1} and choose ${\bf d}_{n,m}=d_n\vec{e}_x$ such that
the {\it strain} $(d_{n+1}-d_{n})/a_0$ is uniformly distributed in
the interval $[-\Delta,\Delta]$ with $\Delta$ the disorder
parameter. The field $B$ sets both the vortex density in the CE's
($\rho_e=(a_0b_0)^{-1}=B/\Phi_0$ with $\Phi_0$ the flux quantum)
and the density $\rho_c=(aw)^{-1}$ {\it inside} the channel.
Hence, vortices {\it in} the channel have an average spacing
$a=a_0b_0/w$ which can be commensurate with the CE arrays
($w=b_0$, $a=a_0$) or incommensurate ($a\neq a_0$). The equation
of motion for vortex $i$ in the channel is:
\begin{equation}
\gamma \partial_t{\bf r} _i = f -\sum_{j\neq i}\nabla V({\bf
r}_i-{\bf r}_j)-\sum_{n,m}\nabla V({\bf r}_i-{\bf R}_{n,m}).
\label{moldyn}
\end{equation}
$V({\bf r})$ is the interaction potential, $j$ labels other
vortices in the channel, $\gamma=B\Phi_0/\rho_f$ with $\rho_f$ the
flux flow resistivity and $f=J\Phi_0$ is the drive along $x$ due
to a uniform current density $J$ applied perpendicular to the
channel.

For $\Delta=0$ the vortex chain can be described by a
Frenkel-Kontorova model for interacting particles in a periodic
potential \cite{RutPRL99}. The ratio between the chain stiffness
and the height of the periodic potential is given by
$g\propto\lambda/a_0\gg 1$ with $\lambda$ the penetration depth. A
commensurate chain depins uniformly at a critical force
$f_c=\mu=2a_0c_{66}/\pi\sqrt{3}$ ($c_{66}$ is the shear modulus)
with a velocity $v=\sqrt{f^2-\mu^2}/\gamma$. At
incommensurability, defects of size $l_d=2\pi a_0\sqrt{g}\gg a_0$
are generated. Their pinning barrier and the critical force are
essentially vanishing. For $f<\mu$ and small defect density, given
by $c_d= |a_0^{-1}-a^{-1}| \lesssim l_d^{-1}$, the motion of
defects leads to a low mobility regime where $v=c_dv_d^0 a_0$ with
$v_d^0=(\pi^2\sqrt{g}/2\gamma) f$ the pure defect velocity
\cite{Landauer}.

We start with a numerical study of the behavior in presence of
disorder. Eq. (\ref{moldyn}) was solved using a modified London
form for $V({\bf r})$ which yields the correct shear modulus
\cite{RutPRL99}. We used cyclic boundary conditions, channels of
length $L\geq 1000a_0$ and we recorded the velocity
$v(f)=\langle\dot{x_i}\rangle_{i,t}$ and vortex positions
$x_i(t)$.

The data points in Fig. \ref{fv} show {\it f-v} curves of a
commensurate and an incommensurate chain for $\Delta=0.025$. We
first focus on the commensurate case. Compared to the result for
ordered CE's (dashed line), $f_c$ in presence of disorder is
clearly reduced. The origin of this reduction is that the random
strains lower the energy barrier for formation of
discommensuration pairs in the chain (interstitial/vacancy pairs
in the 2D crystal formed by the chain and the CE's).

\begin{figure}
\epsfig{file=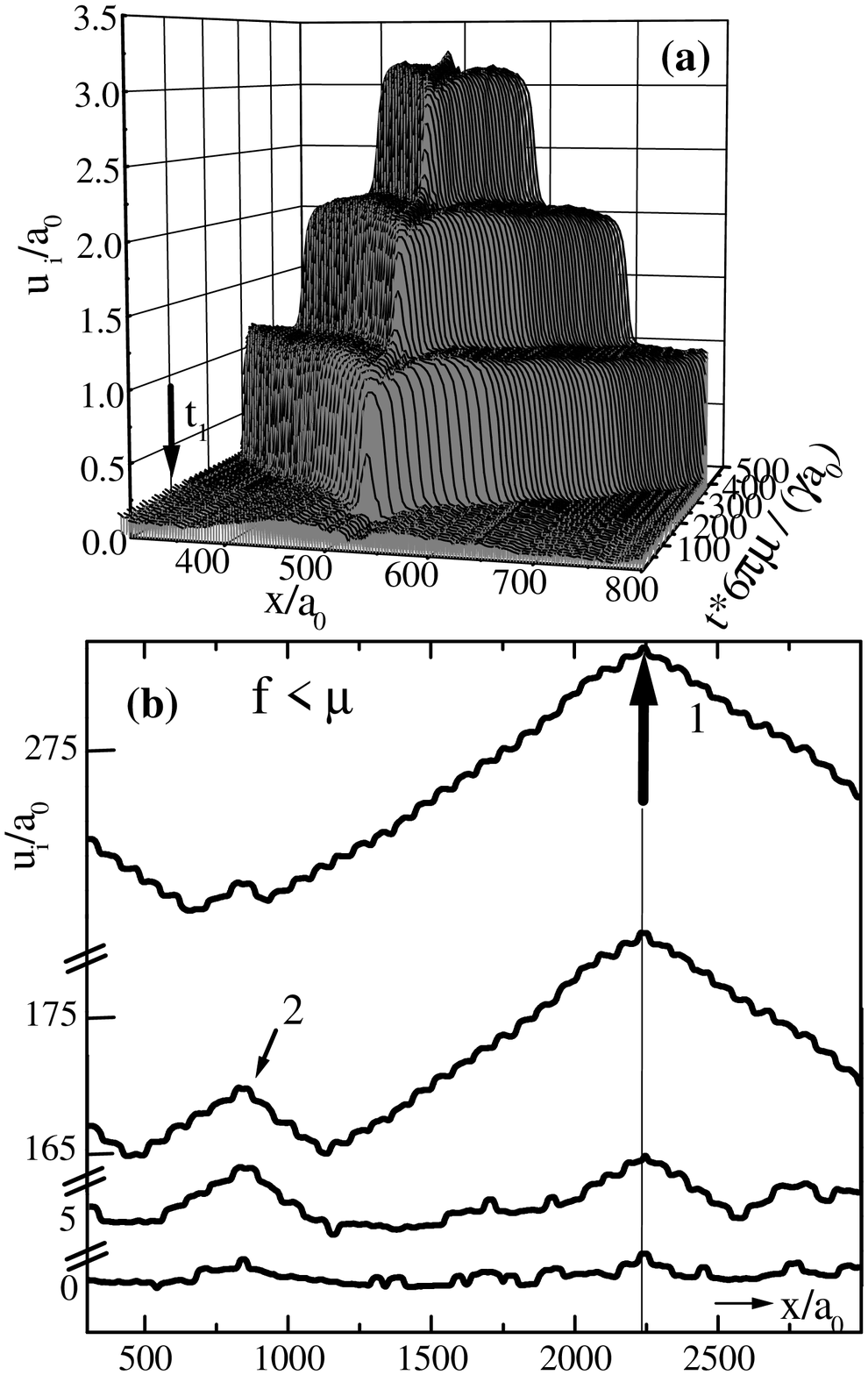, width=6cm, height=8cm,angle=00}
\vspace{0.5cm} \caption{Evolution of {\em longitudinal}
displacements $u_i(t)$ plotted for clarity in a transverse way vs.
$x$ ($\Delta=0.025$): (a) Nucleation of defect pairs in the
commensurate chain of Fig. \ref{fv} ($L=1000a_0$). (b) Transient
response for a system with $L=3000a_0$. The labels 1 and 2 mark
competing nucleation centers in the long time dynamics.}
\label{ufields}
\end{figure}

We show the depinning process in detail in Fig. \ref{ufields}a by
plotting the time evolution of the displacements $u_i=x_i-ia_0$
\cite{website}. At $t=t_1$, the force is increased to a value
$f>f_c$. The motion starts at an unstable site in the chain by
nucleation of a vacancy/interstitial pair shown as steps of $\pm
a_0$ in $u$. We denote the force at which this local nucleation
occurs by $f_n$. The process at this site repeats periodically
with rate $R_n\propto (f-f_n)^\beta$ and a depinning exponent
$\beta=0.46\pm 0.04$, as previously reported for 1D periodic media
\cite{{Myers93}}. Due to the nucleation center, a domain forms
with defect density $c_d=R_n/\langle v_d \rangle$ and a net
velocity $v=c_d \langle v_d \rangle a_0=R_na_0$ with $\langle v_d
\rangle$ the average defect velocity $\simeq v_d^0$, see below. In
a larger system (Fig. \ref{ufields}b), a distribution of unstable
sites $p$ with local rate $R_{n,p}$ initially leads to the growth
of several domains. However, when two domains with rates
$R_{n,1}>R_{n,2}$ meet, their interstitials and vacancies
annihilate and domain 1, with the larger nucleation rate, then
expands at the cost of domain $2$ with a rate $\sim(R_n^1-R_n^2)$.
The stationary state, covering the entire system, is thus governed
by the nucleation center with the smallest local threshold
$f_n^{min}$.

Next, we turn to the {\it f-v} curves in Fig. \ref{fv} at
incommensurability ($w/b_0=0.98$, small defect density
$c_d\sim1/2l_d$). In contrast to the curve for $\Delta=0$ (full
line), the data for $\Delta=0.025$ show a significant threshold
force due to pinning of the existing defects by random strains in
the CE's. We define the pinning force for a single defect as
$f_d(x)$ with a distribution $\{f_d\}$ along the channel and a
maximum value $f_d^{max}$. The shape of the distribution is
roughly gaussian and has a width $\sim 0.1 \mu$. The threshold
force for a single defect will be $f_c=f_d^{max}$. If the defect
density is low, the defects occupy only the highest values of
$\{f_d\}$ and $f_c\lesssim f_d^{max}$ which in Fig. \ref{fv} is
$f_c\simeq 0.2 \mu$. Once defects are depinned, their velocity
$v_d$ fluctuates spatially and $\langle v_d \rangle$ is smaller
than $v_d^0$. However, these effects strongly decrease with
velocity and for $f\gtrsim 2 f_d^{max}$ one retains a low mobility
regime with $dv/df\simeq c_d v_d^0 a_0$, as seen in Fig. \ref{fv}.
Considering the regime of larger drive, the data exhibits a
velocity upturn for $f\sim f_n^{min}$. At this driving force {\em
new} defect pairs start to nucleate at a strong disorder
fluctuation with a rate that exceeds the one at which {\em
existing} defects travel trough the system.

\begin{figure}
\epsfig{file=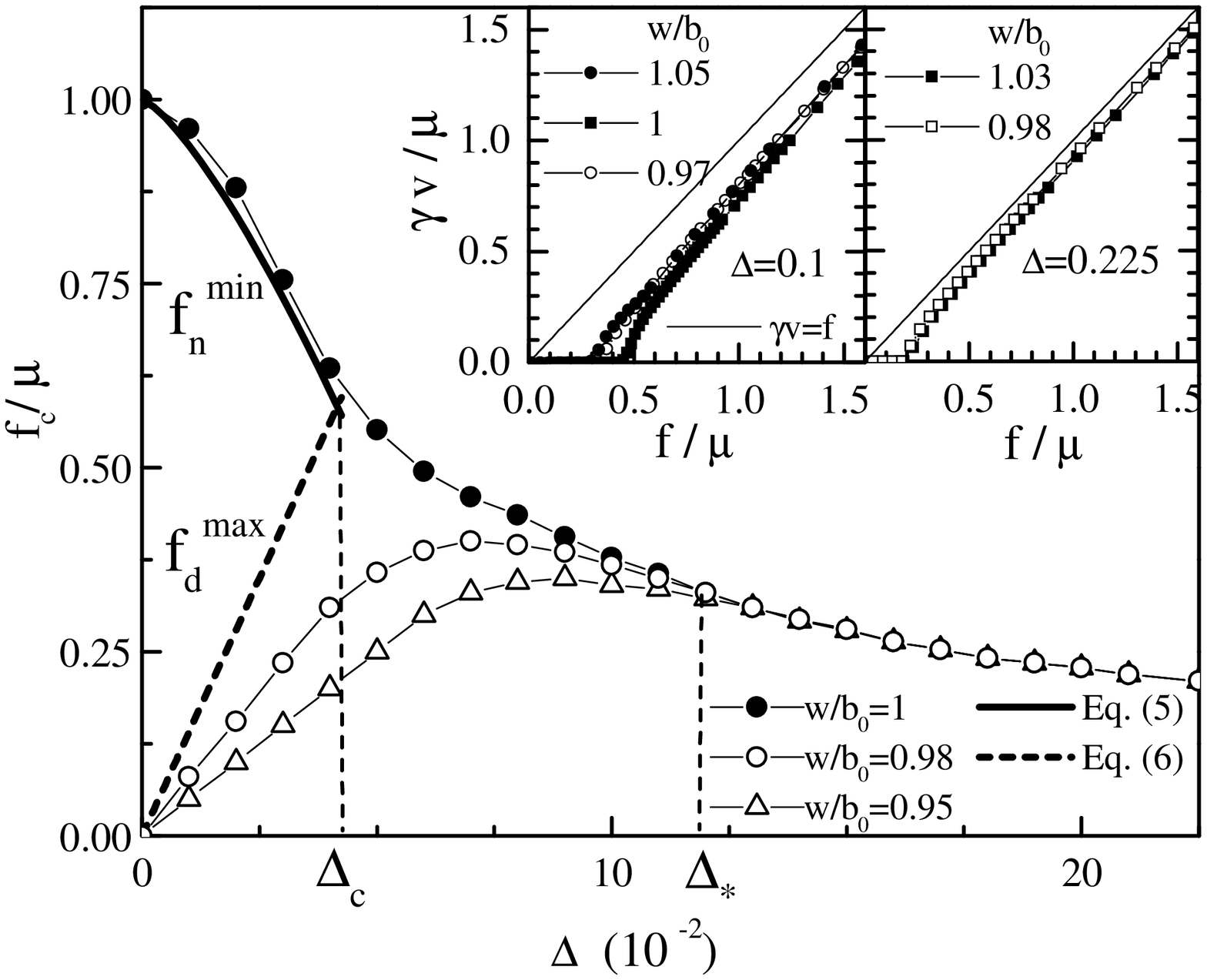, width=7.5cm, height=5.5cm,angle=00}
\vspace{0.5cm}\caption{Threshold force $f_c$, defined by a
criterion $v_c \approx 0.02 \mu/\gamma$, versus $\Delta$ for
$w=b_0$ ($\bullet$), $w/b_0=0.98$ ($\circ$) and $w/b_0=0.95$
($\triangle$). The thick solid and dashed lines represent Eqs.
(\ref{fnmin}) and (\ref{fdmax}). The typical disorder strengths
$\Delta_c$ and $\Delta_*$ are indicated. The insets show {\it f-v}
curves for $\Delta>\Delta_c$.} \label{newphasediag}
\end{figure}

Let us now discuss the disorder dependence of the threshold
forces. Shown in Fig. \ref{newphasediag} are $f_c$ data versus
$\Delta$ for commensurate and incommensurate chains. For $w=b_0$
the minimum from many disorder realizations is plotted, i.e.
$f_c=f_n^{min}$. With growing disorder $f_n^{min}$ decreases
sharply while for the incommensurate chain with $w/b_0=0.98$,
$f_c\sim f_d^{max}$ grows linearly. This behavior changes at
$\Delta \simeq \Delta_c$ where $\Delta_c$ is defined as the
disorder strength where for $w=b_0$ defects first appear
spontaneously. Above $\Delta_c$ both curves (in)decrease more
slowly with disorder and eventually merge. This change in behavior
is due to the fact that favorable nucleation sites in the
commensurate case act as strong pinning site for defects in the
incommensurate case. The simulations show that the presence of
pinned defects strongly affects the formation of a new nucleus. As
a result, $f_n^{min}$ decreases more slowly at larger disorder,
i.e. at a higher density of pinned defects. The curve for
$w/b_0=0.95$, for which the density of static, mismatch induced
defects is $c_d \simeq l_d^{-1}$, shows a reduced threshold force
and merges with the other curves at a disorder strength
$\Delta=\Delta_*$. For $\Delta>\Delta_*$, disorder induced defects
start to overlap (i.e. their density becomes $\gtrsim l_d^{-1}$)
and $f_c$ decreases further. The large disorder also has a
pronounced effect on the shape of the {\it f-v} curves. As shown
in the insets to Fig. \ref{newphasediag}, the typical low mobility
regime at weak mismatch has vanished and all curves exhibit linear
behavior, except in a small regime for $f$ just above $f_c$.

To uncover the underlying physics of the phenomena described
above, we now propose an analytical description of our system. The
energy of a vortex in the channel at ${\bf r_0}=(x,0)$ due to
interaction with shifted vortices in the CE's is:
\begin{eqnarray}
V({\bf r_0})=(2\pi)^{-2}\int d{\bf k}V({\bf k})\rho_e({\bf
k})e^{i{\bf k}\cdot{\bf r_0}}. \label{potential}
\end{eqnarray}
$V({\bf k})=2\pi U_0/({\bf k}^2+\lambda^{-2})$ with
$U_0=\Phi_0^2/2\pi\mu_0\lambda^2$, and $\rho_e({\bf k})$ are the
Fourier transforms of the London potential and the vortex density
in the edge, respectively. For weak disorder ($\nabla \cdot {\bf d
}\ll 1$), $\rho_e$ can be decomposed \cite{GiaPRB95}: $\rho_e({\bf
r}_e,{\bf d})\simeq (B/\Phi_0)(1-\nabla\cdot{\bf d}+\delta
\rho_e)$ where $\delta \rho_e=\sum_i\cos[{\bf K}_i({\bf r}_e-{\bf
d}({\bf r}_e))]$ and ${\bf K}_i$ spans the reciprocal lattice.
Substitution in Eq. (\ref{potential}) yields two terms: a
quasi-periodic potential $V^p\simeq(\mu/k_0)\sin[k_0(x-d)]$, with
$k_0=2\pi/a_0$ \cite{footnote2} due to $\delta \rho_e$ of the
vortex rows nearest to the CE's, and a random, nonlocal
contribution coming from the density fluctuations: $V^r({\bf
r_0})=-(B/\Phi_0)\int d{\bf r}_eV({\bf r_0}-{\bf r}_e)\nabla \cdot
{\bf d}({\bf r}_e)$ with correlator:
\begin{eqnarray}
\langle V^r(0)V^r(x) \rangle \simeq
4\Delta^2U_0^2(\lambda/a_0)^{1+\alpha} e^{-(x/\lambda)^2}.
\label{correlator}
\end{eqnarray}
Here $\alpha$ depends on the disorder correlations between the
rows $m$ in the CE's. We choose $\partial_x d(x)$ to be
independent of row number, which yields $\alpha=2$.

\begin{figure}
\epsfig{file=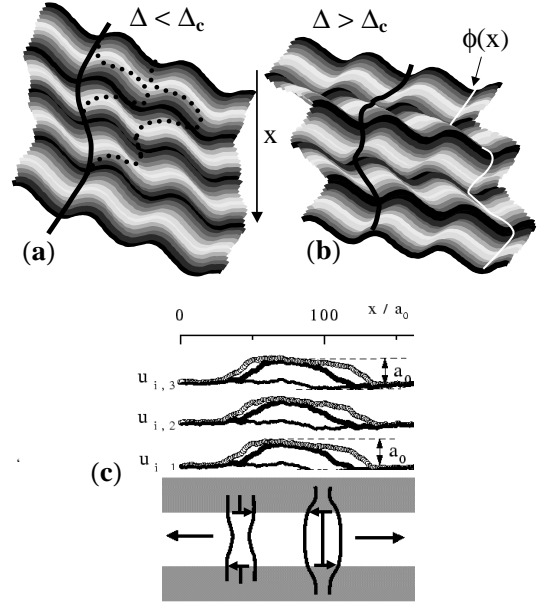, width=7cm, height=8cm,angle=00}
\vspace{0.5cm} \caption{(a,b) Mechanical representation of Eq.
(\ref{contdiseqm}). (a) For $\Delta<\Delta_c$ a gap exists between
the barrier for defect nucleation (dashed) and defect pinning. (b)
When $\Delta>\Delta_c$ disorder induced defects are always present
below $f_c$. The white line shows the random phase $\phi(x)$ of
the washboard potential. (c)top: Evolution of longitudinal
displacements of individual rows at depinning for $w/b_0=3$ and
$\Delta=0.02$. (c)bottom: Square lattice representation of the
nucleated stacks of discommensurations. Small arrows indicate the
Burgers vector of the dislocations terminating each stack.}
\label{washboard}
\end{figure}

To obtain a continuum description of the chain in terms of the
displacement field $u(x)$, the vortex density $\rho_c$ {\em in}
the channel is decomposed, as was done for $\rho_e$. The resulting
interaction of the chain with the CE's is $H_{p}=a_0^{-1}\int(
V^p+V^r)[\delta\rho_c(x,u)-\partial_x u]dx$. For $\lambda> a_0$,
both $V^r$ and $\partial_x u$ vary slowly while $V^p$ and $\delta
\rho_c$ oscillate rapidly and only two terms in $H_p$ remain
\cite{fnrandomphase}. The force $-\delta H_p/\delta u$ on the
chain thus contains a commensurate term and a random compression
term $-\partial_xV^r(x)$, which is independent of $u$. We obtain
the equation of motion for $u$ by adding the {\it intra-chain}
force $\kappa\partial_x^2u$, with stiffness $\kappa\simeq
U_0\pi(\lambda/a_0)$ \cite{footnote3}, resulting in:
\begin{eqnarray}
\gamma\partial_t u=&& f+\kappa\partial_x^2u-\mu\cos(k_{0}u)
-\partial_xV^r(x). \label{contdiseqm}
\end{eqnarray}
The reduced stiffness is now $g=\kappa/(k_0\mu
a_0^2)=3\pi(\lambda/a_0)$. Eq. (\ref{contdiseqm}) describes the
transverse displacements of an elastic string in a tilted
washboard potential with random phase $\phi(x)=-\int_{-\infty}^x
dx'V^r(x')/\kappa$, see Fig. \ref{washboard}. It corresponds to a
commensurate CDW with random forward scattering \cite{Vinokur}
rather than the usual model for a CDW in which the
commensurability potential is ignored either due to strong {\it
direct} random coupling to $u$ or due to large mismatch
\cite{FukuPRB78}.

Next, the dependence of $f_n^{min}$ and $f_d^{max}$ on the strain
$\Delta$ can be addressed by considering a deformation in the CE's
of wavelength $l_{dis}$: $\partial_x d({\bf r})=\Delta\sin(2\pi
x/l_{dis})$. When $l_{dis}\gg \lambda$, the last term in Eq.
(\ref{contdiseqm}) simplifies to $-\partial_xV^r\simeq g^2\Delta
U_0/(2\sqrt{3}l_{dis})\cos(2\pi x/l_{dis})$. To describe
nucleation (for $f\gtrsim \mu/2$), we need the string
displacements $\delta(x)=u-u_0$ relative to the minimum of the
{\em tilted} washboard potential given by
$u_0=-k_0^{-1}[2(1-f/\mu)]^{1/2}$. The restoring force for such
displacements is $\mu k_0^2u_0\delta(x)$, while the elastic force
is $\kappa\partial_x^2 \delta$. Balancing these with
$\partial_xV^r$ yields $\delta(x)=\delta_0\cos(2\pi x/l_{dis})$.
The amplitude of the displacement $\delta_0\sim l_{dis}/(l_d^2+k_0
|u_0| l_{dis}^2)$ is largest for a wavelength
$l_{dis}=l_d/\sqrt{k_0 |u_0|}$ resulting in $\delta_0^{max}\simeq
\sqrt{3}a_0 g^{3/2}\Delta/(8\pi\sqrt{k_0|u_0|})$. Since the {\it
critical} displacement at which the string depins is $\simeq
|u_0|$ \cite{Landauer}, the minimum nucleation force follows from
$|u_0|=\delta_0^{max}$:

\begin{eqnarray}
f_n^{min}/\mu \simeq 1-\left[4\Delta
g^{3/2}/(5\sqrt{3})\right]^{4/3}. \label{fnmin}
\end{eqnarray}

For an {\it existing} defect at zero drive, the pinning energy is
$E_{d,p}(x)=a_0^{-1}\int dx'\partial_x u_d(x'-x)V^r(x')$ with
$u_d$ the familiar shape of a sine-Gordon kink \cite{Landauer}
centered at $x$ and of size $l_d$. Optimal pinning occurs for
deformations in the CE's with $l_{dis}=l_d$, similar to the
'optimal' size of a nucleation center for $f \simeq \mu/2$. The
maximum defect pinning force is then given by:
\begin{eqnarray}
f_d^{max}/\mu \simeq \Delta g^{3/2}/\sqrt{3}. \label{fdmax}
\end{eqnarray}
The results (\ref{fnmin}) and (\ref{fdmax}) are plotted in Fig.
\ref{newphasediag}, using $g=9$ as in the simulations. Eq.
(\ref{fnmin}) agrees with the numerical data, while Eq.
(\ref{fdmax}) can be considered as upper bound. The curves merge
at $\Delta_c \simeq \sqrt{3}g^{-3/2}/2$, where pinned defects can
appear spontaneously in the channel \cite{footnote4}.

The disorder strength $\Delta_*$ where the density of disorder
induced defects becomes $\sim l_d^{-1}$ can be estimated by
equating the typical (rms) pin energy of a defect with its bare
energy $\sim \mu a_0\sqrt{g}$. The former is estimated using the
previous form for $E_{d,p}$ and Eq. (\ref{correlator}): $\langle
(E_{d,p})^2 \rangle^{1/2}\simeq
2U_0(g/3\pi)^{(2+\alpha)/2}g^{-1/4}\Delta$ leading to
\begin{eqnarray}
\Delta_*\simeq (3\pi/4) g^{-5/4}, \label{deltastar}
\end{eqnarray}
also in reasonable agreement with the data in Fig.
\ref{newphasediag}. These results show that the effect of disorder
rapidly grows on increasing the interaction range $\lambda/a_0$.

Finally, we shortly discuss how these results carry over to
channels with {\em multiple} chains near matching ($w/b_0\simeq
n$, with $n$ an integer $\geq 2$). Without disorder $f_c$ has
sharp peaks at matching of height $f_c^0=\mu b_0/w=\mu/n$
\cite{RutPRL99}. With disorder, the behavior of $f_c$ is similar
to that in Fig. \ref{newphasediag} \cite{tobe}. In Fig.
\ref{washboard}(c) we show the depinning process for $w/b_0=3$,
$\Delta=0.02$ and $f=0.7 f_c^0$. As observed, the defects involved
with depinning consist of stacks of discommensurations, coupled
between the chains. Each stack is terminated by a pair of
dislocations with opposite Burgers vector {\em along} the CE's.
This quasi-1D behavior at weak disorder can also be described by
Eq. \ref{contdiseqm} by substituting $\mu/n$ for $\mu$. For larger
disorder the threshold near matching reduces to $\sim 30 \%$ of
$f_c^0$. The merging of the commensurate and weakly incommensurate
threshold forces now occurs for $\Delta_*^n=\Delta_*/n^{1/4}$. For
a realistic value of $\lambda/a_0 \sim 4$, we estimate $\Delta_*
\sim 0.03$ meaning that rms strains in the CE's $\lesssim 2 \%$
already cause strong disorder. At strong disorder {\em and large
mismatch} the quasi-2D nature of the system causes new phenomena:
for $w/b_0\simeq n \pm 1/2$ transitions from $n\rightarrow n\pm 1$
rows occur, involving pinning of dislocations with {\em
misaligned} Burgers vector \cite{KokuboPRL02}. The threshold force
then {\it exceeds} $f_c$ near matching \cite{KokuboPRL02}.

In summary, depinning of a vortex chain in a channel formed by
disordered vortex arrays with (nearly) the same periodicity,
occurs by nucleation of defect pairs or motion of existing
defects. The gap between the barriers for these two processes and
the sharp peak in $f_c$ at commensurability vanish with increasing
disorder. For large disorder in the channel edges $f_c$ saturates
at a small fraction of the ideal lattice strength.

This project is supported by the Nederlandse Stichting voor
Fundamenteel Onderzoek der Materie (FOM) and also by U.S.DOE,
Office of Science under contract $\#$W-31-109-ENG-38 during R.B.´s
stay at Argonne. V.M.V. is supported by U.S.DOE, Office of Science
under contract $\#$W-31-109-ENG-38.

\end{multicols}

%\clearpage
%\newpage
%\ FIGURE CAPTIONS

\begin{references}
\bibitem{fnrandomphase}The product $V^p\rho_c$ in fact yields
$(\mu/k_0) \sin[k_0(u-d)]$. The resulting extra random phase
$-k_0d$ in the $\cos$ term in (\ref{contdiseqm}) is ignored since
it is small compared to the phase $\phi$ arising from $V^r$.
\bibitem{footnote3}The full, dispersive expression for the stiffness is
$\kappa(k)=U_0\pi(\lambda/a_0)/\sqrt{1+\lambda^2k^2}$. Thus Eq.
(\ref{contdiseqm}) applies for $k<\lambda^{-1}$.
\bibitem{footnote4}Since, for depinning, the 'optimal' wavelengths are at least $l_d$,
the results are valid when $l_d>2\pi\lambda$. Using $l_d=2\pi
a_0\sqrt{g}$, this corresponds to the range $\lambda/a_0 \lesssim
10$.
\end{references}
\end{document}